\documentclass{iaus}
\usepackage{graphicx}
\title[Effects of the Environment on the Properties of Seyfert Galaxies]{Effects of the 
Environment on the Properties of Seyfert Galaxies}
\author[M.A.G. Maia et al.]{Marcio A. G. Maia$^{(1,2)}$, Christopher N. A. Willmer$^{(2,3)}$, Bruno M.
 Rossetto$^{(1)}$, Rodolfo S. Machado$^{(1)}$}
\affiliation{$1$ Observat\'orio do Valongo, Universidade Federal do Rio de Janeiro, Ladeira do
Pedro Ant\^onio 43, Rio de Janeiro - 20080-090, RJ, Brazil
\\[\affilskip] $^2$ Observat\'orio Nacional, Rua Gal. Jos\'e Cristino 77, Rio de Janeiro -
20921-900, RJ, Brazil 
\\[\affilskip] $^3$ Lick Observatory, University of California, Santa Cruz, 1156 High Street, 
Santa Cruz, CA 95064, USA}

\pubyear{2004}
\volume{222}
\pagerange{1--8}
\date{?? and in revised form ??}
\jname{The Interplay among Black Holes, Stars and ISM \\in Galactic Nuclei}
\editors{Th. Storchi Bergmann, L.C. Ho \& H.R. Schmitt, eds.}
\begin{document}
                                                                                
\maketitle

\begin{abstract}
We identify 175 Seyfert galaxies from the Southern Sky Redshift Survey
- SSRS2. We use the entire SSRS2 catalog to investigate the correlation
between the presence of AGN with host environment. The AGN phenomenon is
more strongly correlated with the internal host properties, than with the external
environment. In particular, we find that Seyferts reside in more luminous galaxies,
and are twice as frequent in barred galaxies and systems showing sign of advanced 
merger condition, when compared to a control sample.
\end{abstract}

\section{Introduction and Sample Definition}
In order to study the dependence of AGN properties with local host
environment and to derive reliable statistical results (see
contribution of H. Schmitt in these proceedings), it is extremely
important to select a suitable control sample. In this work we
use the SSRS2 catalog \cite{daC98}, which contains 5399 galaxies with $m_B \le 15.5$ 
covering 1.69 sr of the southern sky. 
This catalog is currently 99.99\% complete in redshift, and in our 
database we have optical spectra available for $\sim$ 78\% of the SSRS2 galaxies, while 
the remaining radial velocities were obtained from the literature. The spectra were 
inspected and AGNs identified using the diagnostic diagrams of line intensity ratios 
proposed by \cite{Bal81}. New Seyfert galaxies were detected 
during the survey and were reported in the papers by Maia et al. (1987
and 1996). Additional AGN hosts in the SSRS2 were obtained inspecting the 
NASA Extragalactic Database, and the literature. For AGN host galaxies without
spectra in the SSRS2 database, new observations were taken, therefore
keeping the identification process homogeneous. Our AGN sample is made
up of 175 Seyferts, which comprise ~3-4\% of  
the parent sample. The ratio Sey-2/Sey-1 is 3:1. This fraction may contain an excess of 
Sey-2 which are in reality Sey-1 with broad lines that are partially
obscured by the material in the disk of host galaxies seen almost edge-on.

\section{Internal Properties of Seyfert Hosts}
\begin{figure}
\center
  \includegraphics*[angle=270, scale=0.45]{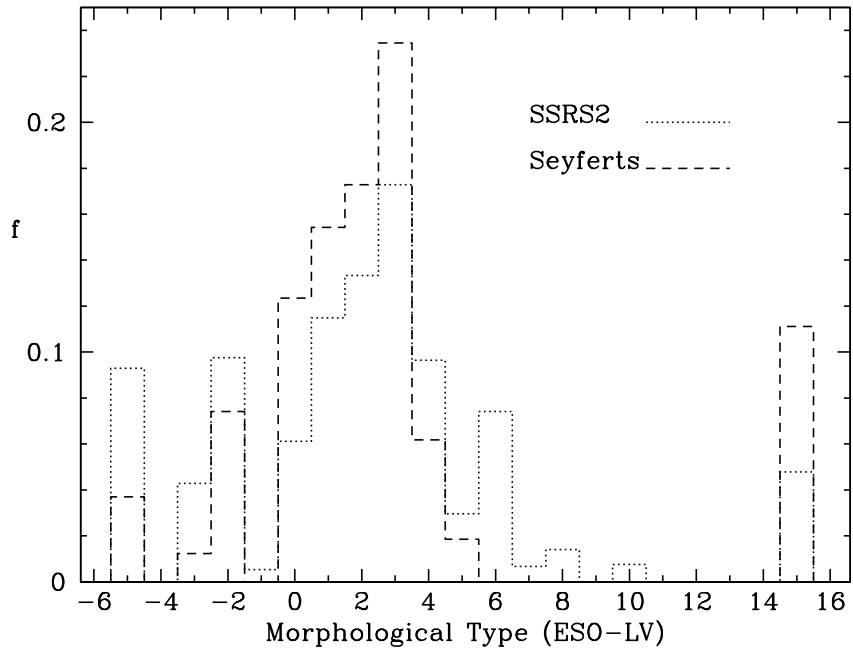}
 \caption{Morphological distribution of Seyferts and SSRS2 galaxies. The numbers correspond 
   to the following classification: E=-5, E/S0=-3, S0=-2, S0a=0, Sa=1, Sab=2, Sb=3, Sbc=4, 
   S...=5, Sc=6, Sc/Irr=7, Sd=8, Irr=10, merger=15.}
 \label{fig:f1}
\end{figure}

We first examine the distribution of morphologies of the Seyfert galaxy population, which 
is displayed in figure~\ref{fig:f1}. Also in this figure is shown the morphological distribution 
of the SSRS2 parent catalog.
Seyfert galaxies are distributed preferentially among the S0a-Sb interval and in those 
hosts showing strong evidence of mergers (type 15). In particular, the excess of Seyferts in 
type 15, could be an indication that strong interactions may trigger or enhance the nuclear 
activity. No significant difference was seen between the host morphologies
of Seyfert types 1 and 2.

\begin{figure}
\center
\includegraphics*[angle=270, scale=0.4]{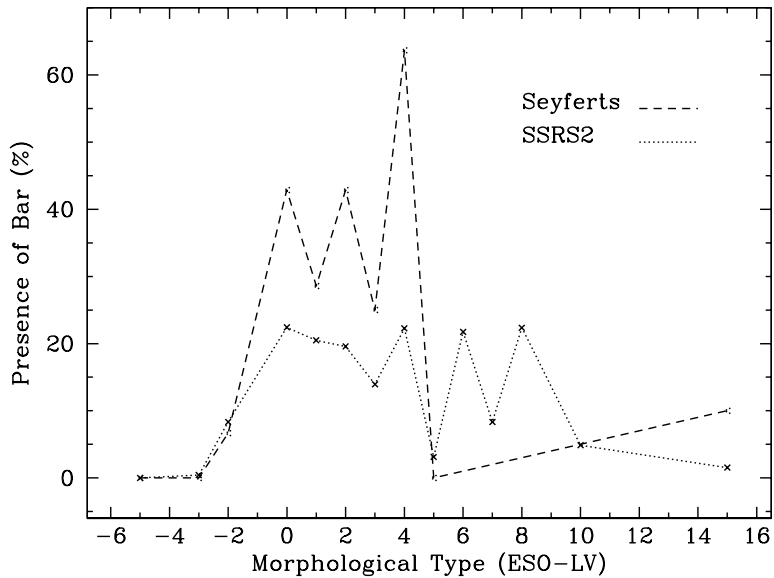}
 \caption{Fraction of galaxies with bars in the Seyfert and SSRS2 samples as a 
         function of morphological type, using the same coding as in figure~\ref{fig:f1}.}
 \label{fig:f2}
\end{figure}

We also investigate the presence of bars, which have been claimed as a
possible mechanism to induce gas and dust radial inflows, fueling the
AGN activity. We inspected visually DSS images of the entire SSRS2
catalog members. Because of saturation in the central parts of
galaxy images in the DSS, the presence of bars cannot be identified
for many galaxies, while for galaxies that are edge-on or which have small
apparent sizes bars are also hard to recognize. Nevertheless, we are
able to detect, at least, the relative differences 
between the Seyferts and control samples. Globally, we find that the
SSRS2 contains 14\%  barred galaxies, while the Seyfert sample
contains 28\%. The fraction of barred galaxies 
as a function of morphological type, for the SSRS2 and Seyfert samples, is displayed in 
figure~\ref{fig:f2}. There are twice as many barred hosts in the Seyfert
population than in the SSRS2 control sample. This result is maintained
even if we consider only the morphological types 
where Seyferts are more frequent. This fraction is the same for subsamples limited at
different distances. 

We examined the total luminosity distributions of the 
AGNs and SSRS2 samples. We find that Seyferts are prone to reside in more luminous hosts, 
when the ratio between the number densities per magnitude bin of Seyfert hosts and SSRS2 
objects are computed.

\section{The External Environment of Seyfert Galaxies}
We examine the possible correlation between the presence of AGN phenomenon with the local density 
of galaxies by identifying  groups of galaxies in the SSRS2. These groups are generated using a 
percolation algorithm analogous to that described by \cite{Mer00}. The algorithm also outputs a 
list of objects for which no companions are found ("isolated galaxies"). 
The assignments for Seyfert and SSRS2 samples according to different degrees of multiplicity 
are displayed in table~\ref{tab:t1}. Both samples are distributed in a similar way, except for binary 
systems, in which Seyfert hosts are more commonly found when compared to other galaxies of the 
entire SSRS2. This suggests that galaxies undergoing close interactions could have the AGN 
phenomenon intensified.

\begin{table}
  \caption{SSRS2 and Seyfert galaxies versus environment assignment}
  \smallskip
    \begin{center}
      {\small %\scriptsize
      \begin{tabular}{lcccc}
%      \tableline
      \noalign{\smallskip}
      Galaxies & Groups & Triplets & Binaries & Isolated\\
               & (\%)   & (\%)     & (\%)     & (\%)    \\
      \noalign{\smallskip}
%      \tableline
      \noalign{\smallskip}
      Seyferts & 21 & 7 & 28 & 44\\
      SSRS2    & 24 & 8 & 18 & 50\\
      \noalign{\smallskip}
%      \tableline
      \end{tabular}
      \label{tab:t1}
      }
    \end{center}
\end{table}

\begin{figure}
\center
\includegraphics*[angle=270, scale=0.4]{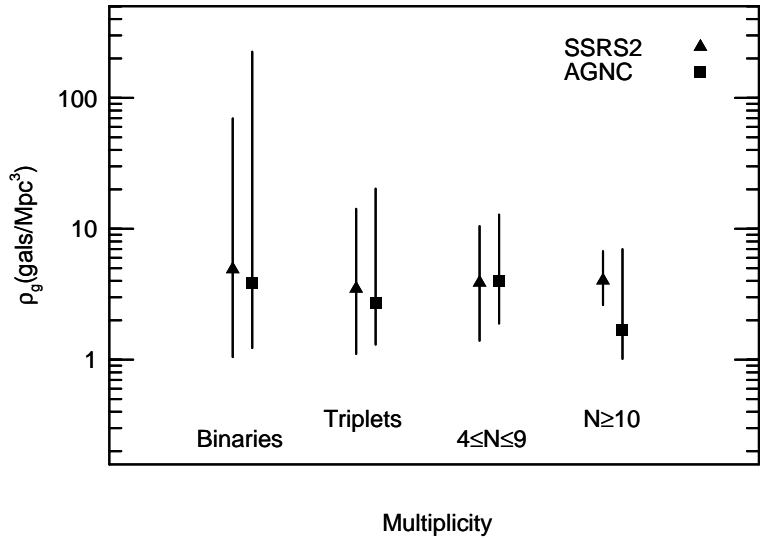}
 \caption{Group density $\rho_g$, according to multiplicity intervals. The points represent
          median values, while the bars refer to upper and lower quartiles of the distributions.}
 \label{fig:f3}
\end{figure}

We calculate the average group density, $\rho_g$, for both the AGN and SSRS2, by means of the
expression $\rho_g = 3 N_g / 4 \pi r^3$ where, $N_g$ is the number of galaxies and $r$ the 
mean pairwise separation of a given group. In figure~\ref{fig:f3} we display the median values 
for $\rho_g$ as a function of group multiplicity. There is no evidence of the AGN phenomenon 
to be correlated with group density.

Some results in the literature show evidence that the presence of a close companion may be 
correlated to the AGN activity. We tested this hypothesis by computing the nearest neighbor 
distance - $S$, and the maximum tidal influence - $Q$ that a companion may exert.  
The tidal influence is proportional to the companion's mass divided by the cube of the 
separation, $S$.  Assuming that light traces mass, the tidal parameter is $Q \propto L/S^3$. 
The distribution of the nearest neighbor separation, $S$, for Seyferts
and SSRS2 galaxies was examined using the KS test which shows that
both samples are similar at the 98\%  
confidence level. The distribution of $Q$ values for the AGNs and SSRS2 are identical at 
the 86\% confidence level. Meanwhile, the Seyfert types 1 and 2 distributions are considered 
the same only at the 20\% confidence level, with type 1s presenting higher values of $Q$.

\section{Conclusions}
We used the magnitude-limited SSRS2 to identify Seyfert galaxies, and investigated how the 
properties of their hosts relate to general population of non-active galaxies.
We find that 175/5339 (3-4\%) of the galaxies in the SSRS2 are Seyferts. The ratio of 
Seyfert 2s to 1s is $\sim$3:1. Most Seyferts are of morphological types 
betwen S0a and Sb, and $\sim$ 10\% have hosts with indication of an ongoing 
merger. There is no difference between the morphology of Seyfert types 1 and 2. The AGNs 
are preferentially detected in high luminosity hosts, and are twice as frequent in barred 
hosts. No correlation with local density of galaxies. We find marginal evidence that 
Seyfert 1 galaxies have closer companions and/or are more susceptible to tidal effects than 
type 2. Additional details about the results presented here can be obtained in the paper by 
\cite{Mai03}.

%\acknowledgements{RSM acknowledges financial support from PIBIC-CNPq scholarship, MAGM to 
%CNPq grant 301366/86-1 and FAPERJ E-26/171.619/2001, and CNAW to NSF AST 95-29028 and 
%AST 00-71198.}

\end{document}